\documentclass[amsfonts,amssymb,amsmath,prl,twocolumn]{revtex4}

\usepackage{times}
\usepackage{graphicx}
\usepackage{bm}
\usepackage{changes}

\usepackage{color}

\definecolor{pink}{rgb}{1,1,0} 
\definecolor{red}{rgb}{1,0,0}
\definecolor{yellow}{rgb}{1,1,0}
\definecolor{orange}{rgb}{1,0.5,0}
\definecolor{white}{rgb}{1,1,1}
\definecolor{blue}{rgb}{0,0,1}

\begin{document}

\title{Finite-time localized singularities as a mechanism for turbulent dissipation.}
\author{Christophe Josserand\textsuperscript{1},  Yves Pomeau\textsuperscript{1}, Sergio Rica\textsuperscript{1,2} \thanks{sergio.rica@uai.cl}}
\affiliation{ \textsuperscript{1} LadHyX, CNRS \& Ecole Polytechnique, UMR 7646, IP Paris, 91128, Palaiseau, France.\\
\textsuperscript{2} Facultad de Ingenier\'ia y Ciencias and UAI Physics Center, Universidad Adolfo Ib\'a\~nez, Avda. Diagonal las Torres 2640, Pe\~nalol\'en, Santiago, Chile.
}

\begin{abstract}
We provide a scenario for a singularity-mediated turbulence based on the self-focusing non-linear Schr\"odinger equation, for which sufficiently smooth initial states leads to blow-up in finite time. Here, by adding dissipation, these singularities are regularized, and the inclusion of an external forcing results in a chaotic fluctuating state. The strong events appear randomly in space and time, making the dissipation rate highly fluctuating.  The model shows that: i) dissipation takes place near the singularities only, ii) such intense events are random in space and time, iii) the mean dissipation rate is almost constant as the viscosity varies, and iv) the observation of an Obukhov-Kolmogorov spectrum with a power law dependence together with an intermittent behavior using structure functions correlations, in close correspondence with fluid turbulence.
\end{abstract}

\maketitle

{\it Introduction.--}  
One of the long-standing problems of classical physics is a thorough understanding of fully developed turbulence. More specifically there is still no explanation grounded on properties of the fluid equations for the observed intermittency in flows at very large Reynolds number. This phenomenon displays strong fluctuations of the velocity and related fields~\cite{BatchelorTownsend49,Sreenivasan87,PuBo19}, fluctuations that seem to preclude any theoretical description based on mean-field theory, even a modified one \cite{frisch}. A possible scenario is that intermittencies are consequence of spatio-temporal singularities of the incompressible Euler equations~\cite{yves19,martine19}.  The addition of dissipation, {\it i.e.} viscosity leading to the Navier-Stokes equation (NS), regularizes these hypothetic singularities, resulting in a chaotic state and making the energy dissipation rate a highly fluctuating quantity. 
The problem of the existence of singular solutions of Euler equations is still unresolved and traces back, at least, to the 1920s~\cite{Lichtenstein,Gunther}. In 1934, Leray questioned the existence of finite-time singularities for the velocity field as point-like singularities in space-time~\cite{Leray}. 
The existence of such singularities regained interest in the eighties and later thanks to the improvement of computers but without yet an unambiguous result~\cite{Siggia85,PumirSiggia87,Brachet92,Gibbon08,Eggers18}. 
Beside turbulence, dissipation by singularity events has been proposed in different physical contexts: for instance, it is believed that the formation of steep slope deformations of the sea surface, leading to ``white caps", are responsible for energy dissipation \cite{NewellZakharov,Pomeau2011}. Ridges, folds and conical singularities are good candidates for dissipating the energy of strongly vibrating elastic plates \cite{DuringPRF} and the same is true for the focusing of light in nonlinear media \cite{Dyachenko-92} and strong turbulence in plasmas \cite{Goldman84}. 

 The present paper aims at studying the possible role of Leray-type finite-time singularities in a simpler model of turbulence where singularities are present in the inviscid limit in order to shed light on their role on the turbulent behavior. 
 The motivation is, therefore, to circumvent the difficulty posed by the solutions of the incompressible Euler/Navier-Stokes equations that makes their thorough analytical studies hard, at least for the moment. 
Over the years, many different models have been suggested
such as the Burgers \cite{Bec07}, the Kuramoto-Sivashinsky \cite{Manne81,Pumir84}, or the complex Ginzburg-Landau (CGL)\cite{Bartuccelli1989,Bartuccelli1990} equations, leading to specific turbulent characterizations. Our model, based on the focusing non-linear Schr\"odinger (NLS) equation (and similar to some extent to the CGL equation), exhibits well-understood singularities in the inviscid limit, so that one can investigate numerically thoroughly how such structures are related to the large scale dynamics.

{\it Theoretical model.--}
We thus consider the focusing NLS equation with small damping and small forcing:
   \begin{eqnarray}
i  \frac{\partial \psi}{\partial t}  &=& - \frac{\alpha}{2}   {\bm  \nabla}^2 \psi  -g  |\psi|^{2n} \psi  -  i \nu { \Delta } ^2   \psi + f_{k_0}({\bm x},t).
\label{eq:NLS}\end{eqnarray}
Here $\psi({\bm x},t)$ is a complex field defined in an infinite space of dimension $D$, the parameter $\alpha>0$ quantifies the dispersion and $g>0$ the nonlinear strength (for the following, we do not consider here the defocusing case which is well known to manifest a quantum vortex-like turbulence \cite{Nore97,Brachet19}). 
Without ``viscosity'' $\nu$ and forcing $f_{k_0}$, one recovers the NLS equation which is conservative and reversible. Indeed, taking, for instance, a periodic box, the quantities 
 \begin{eqnarray}
N&= & \int  |\psi|^2   d^Dx, \quad {\rm and}\label{eq:Masse}\\
H &= & \int \left(  \frac{\alpha}{2 }  |{\bm \nabla} \psi|^2   - \frac{g}{n+1} |\psi|^{2 (n+1)} \right) d^Dx, \label{eq:Energy}
\label{energy} 
\end{eqnarray}
are conserved in time. Notice that while the ``mass" $N$ is positive, the sign of the ``energy" $H$ is not prescribed.

\begin{figure}[h!]
 \centering 
(a) \includegraphics[width=0.2\textwidth]{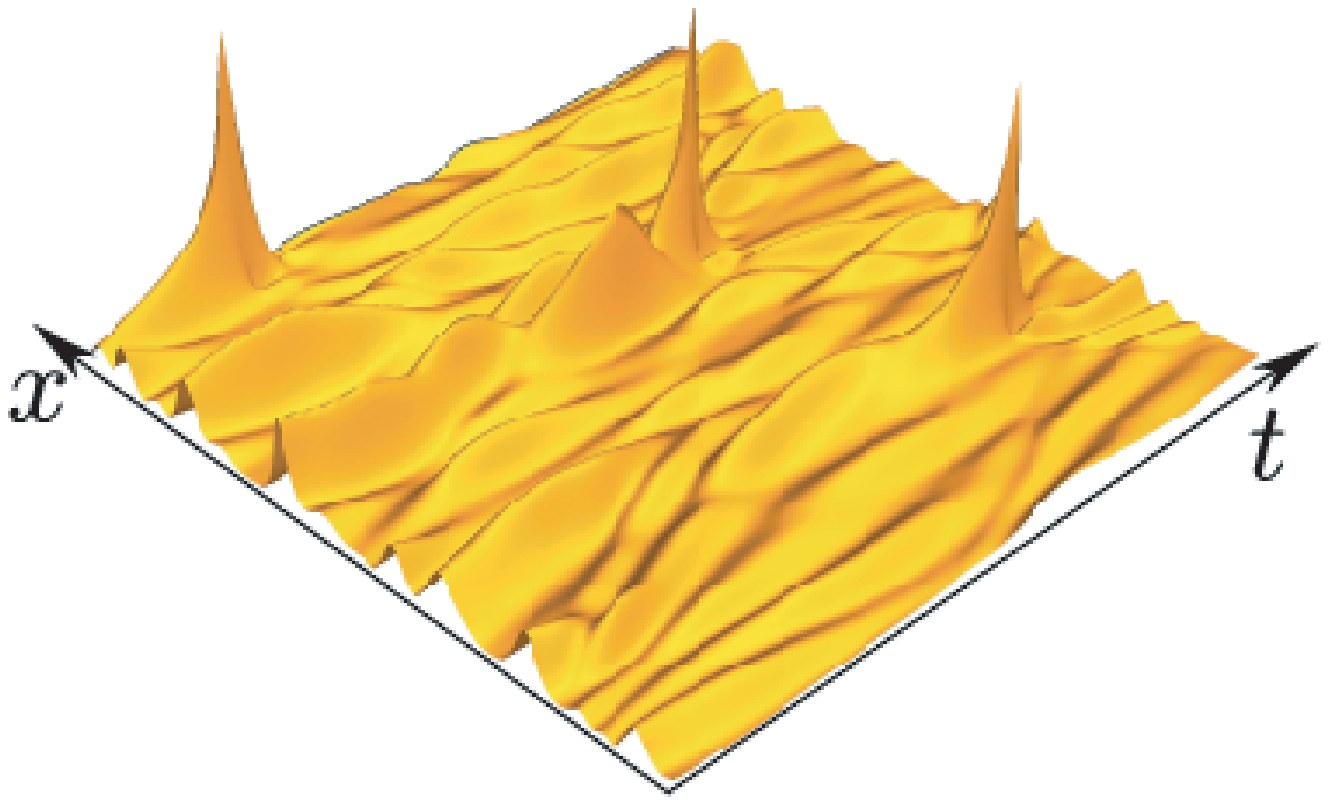}(b)
\includegraphics[width=0.2\textwidth]{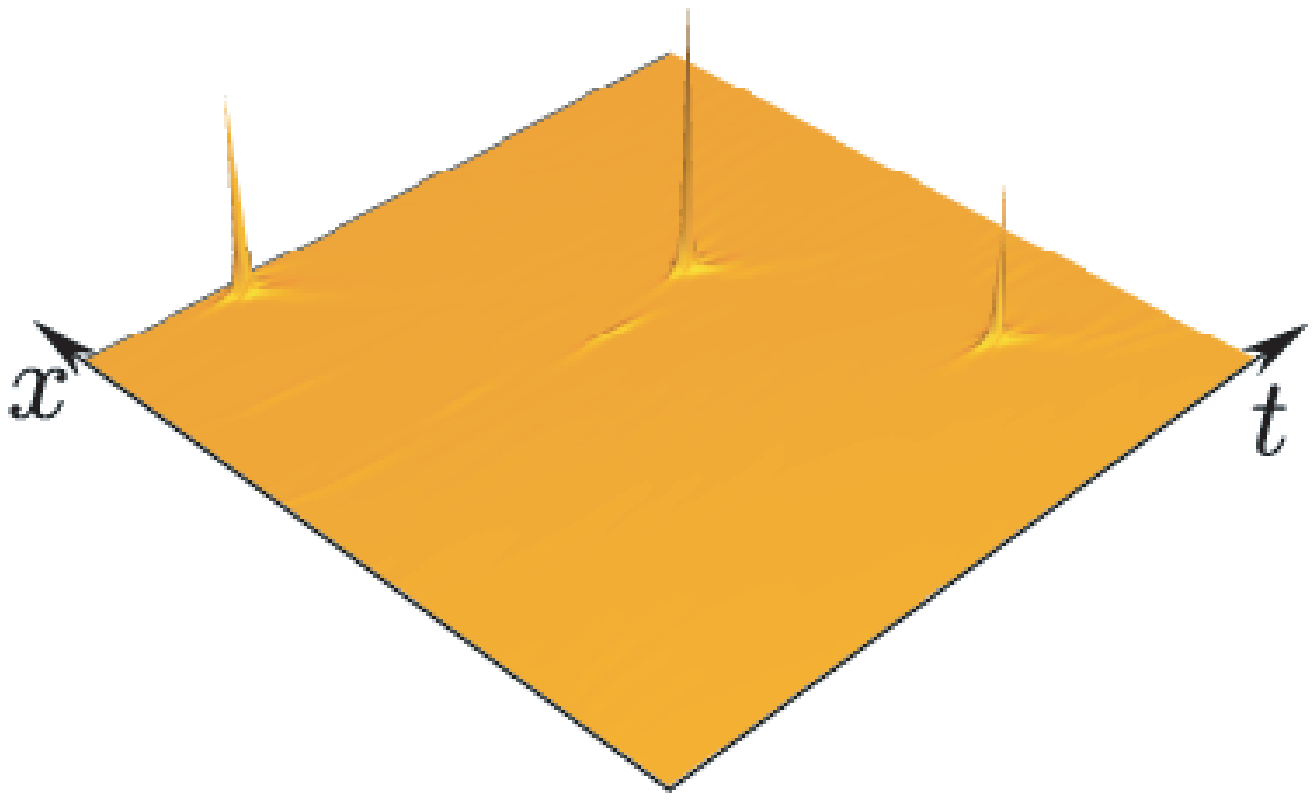} 
                   \caption{ 
$x-t$ diagram for numerical simulations of  (\ref{eq:NLS}) for $D=1$ and $n=3$. (a) $|\psi(x,t)|^2$ and  (b) the local and instantaneous dissipation $ \epsilon(x,t)$. Remarkably, as expected, the dissipation is strong near the large events displayed by $|\psi|^2$ and negligible otherwise.  As a scale reference, the maximum value of ${\rm max}\, \{ |\psi(x,t)|^2\} \approx 6.67$, while for the dissipation  ${\rm max}\, \{ \epsilon(x,t) \} \approx 600$.  
The simulations were done using pseudo-spectral method with $1024$ modes, a mesh size of $dx=0.05$ (so that the computational periodic domain is of size $L=51.2$), $dt =  10^{-6}$,  $\nu  = 2.5 \times10^{-5}$ and  $g=\alpha = 1$. Finally, the random forcing acts for wave-numbers $k<k_0=0.3$ with a forcing amplitude $a = 0.01$.}
  \label{fig:xtDiagram}
\end{figure}

The term $-  i \nu { \Delta } ^2   \psi $ in (\ref{eq:NLS}) denotes damping which we have chosen to bear two orders of derivation higher than the inviscid and conservative case ($\nu=0$), by contrast with the usual CGL equation investigated earlier \cite{Bartuccelli1989,Bartuccelli1990}. Finally, $f_{k_0}({\bm x},t)$ is an added complex forcing acting at large scales of order $\sim 1/k_0$ (called the integral scale hereafter). Equation (\ref{eq:NLS}) is complemented with a smooth initial condition. Fig. \ref{fig:xtDiagram}( a) shows a typical $x-t$ diagram of $|\psi(x,t)|^2$ obtained by numerical simulations of eq. (\ref{eq:NLS}) performed for $D=1$, $n=3$: a turbulent regime is observed with large amplitude events localized in space and time. These events display strong gradients and, more importantly, correspond to very strong local dissipation (defined below as $\epsilon(x,t)= 2 \nu | \partial_{xx}\psi|^2$,  see eq. (\ref{eq:Ndissipation})), as displayed on Fig. \ref{fig:xtDiagram} (b).

Indeed, one exciting feature of the model (\ref{eq:NLS}) is that in this conservative limit ($f_{k_0}=0$ and $\nu=0$), the solutions of (\ref{eq:NLS}) may display a finite-time singularity at a given point (a position and a time hard to predict). 
More precisely, if $ 2< n D < 2(n+1)$ and for smooth initial conditions such that initially $H< 0$, then, the solution of (\ref{eq:NLS}) blows-up at a point (the solution and its gradient become infinite) in finite-time. Moreover, the singularity formation is self-similar so that its amplitude increases while the size of the blow-up region decreases \cite{sulem,budd}. In our model, eq. (\ref{eq:NLS}), this singularity is avoided thanks to the viscous term and a turbulent regime appears where 
forcing and dissipation balance each other.

Indeed, in the absence of forcing, $f_{k_0}=0$, the model  (\ref{eq:NLS}) admits uniform solutions $\psi_0({\bm x},t) = \sqrt{\varrho_0} e^{i g \varrho_0^{n}t} $,  where $\sqrt{\varrho_0} $ is constant, whose linear stability analysis leads to the
following dispersion relation (considering the perturbed solution $\psi({\bm x},t) =  \psi_0({\bm x},t)  + \delta \psi({\bm x},t)  e^{ig \varrho_0^{n} t}$):
 \begin{eqnarray}
\sigma^{(\pm)}_k &=&   - \nu k^4 \pm  \sqrt{ \alpha n  g \varrho_0^{n}  k^2   - \frac{\alpha^2 k^4}{4}  } .
\label{eq:Bogoliubov}\end{eqnarray}

In the long wave limit one of these eigenvalues, namely $\sigma^{(+)}_k \approx |k|  \sqrt{ g\alpha n \varrho_0^{n} } $,  develops a long wave modulation instability, while short wavelength perturbations (for $k \geq 2 \sqrt{ng  \varrho_0^{n}/\alpha}$) are stable and propagate dispersively. 
For all $\nu>0$ the uniform solution is still linearly unstable for a well-defined bandwidth and for any smooth initial condition this modulation instability can trigger a self-focusing mechanism for wave collapse.  As the solution becomes narrow and broad in amplitude, the viscous dissipation becomes more critical
and, eventually, is expected to avoid the inviscid blow-up. This instability mechanism can be seen in numerical simulations 
 by following the time evolution of the equation for different viscosities $\nu$ starting from the same smooth bump initial condition. Although the model exhibits singularities for any space dimensions, we will focus later on the 1D version for the sake of simplicity, since it displays the essence of the dynamics and allows for high accuracy numerical simulations ($D=1$ and $n=3$ for all the numerical results shown below). 
Fig. \ref{fig:Turbulence} (a) plots the maxima of the squared modulus of the wave-function $|\psi_0(t)|^2={\rm max}_x (|\psi(x,t)|^2)$ as functions of time for different viscosities $\nu$: we observe that the smaller the viscosity, the sooner and the higher is the peak of $|\psi_0(t)|^2$. This indicates that the viscosity indeed cures the singularity of the inviscid equation in such a way that the peak intensity diverges as the viscosity vanishes. 
 Moreover, defining the instantaneous Fourier spectrum $S_k(t)$ by
 \begin{eqnarray}S_k(t) &\equiv& |\hat \psi_k|^2+ |\hat \psi_{-k}|^2,
 \label{eq:SpectrumDef}
\end{eqnarray}
where the Fourier transform reads $\hat \psi_{ k} (t) = \frac{1}{L^{1/2}} \int \psi({ x},t) e^{i { k} { x} } d { x}$, we observe that the spectrum just before the peak and the one on the peak show a smooth behavior compatible with the scaling law $k^{-4/3}$ that can be deduced from the self-similar singularity solution (see Supplemental Material \cite{SupplMat}). 
On the other hand, the spectrum for the time step just after the peak exhibits a different behavior with the rapid formation of small scales (large $k$) fluctuations, witnessing the oscillations observed in the dynamics after the peaks of Fig. \ref{fig:xtDiagram}.

 \begin{figure}[h!]
 \centering 

(a)\includegraphics[width=0.21\textwidth]{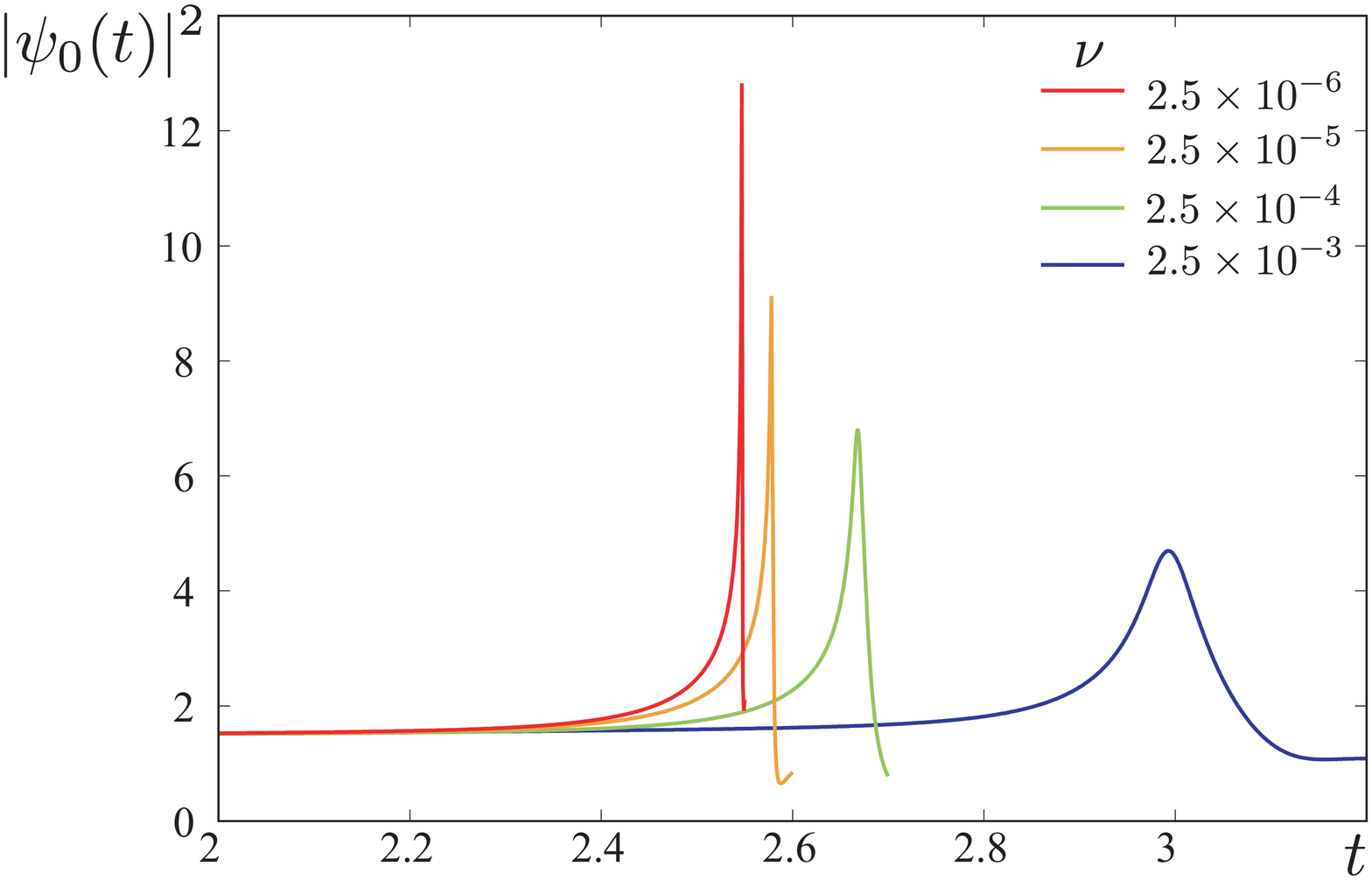} (b)\includegraphics[width=0.185\textwidth]{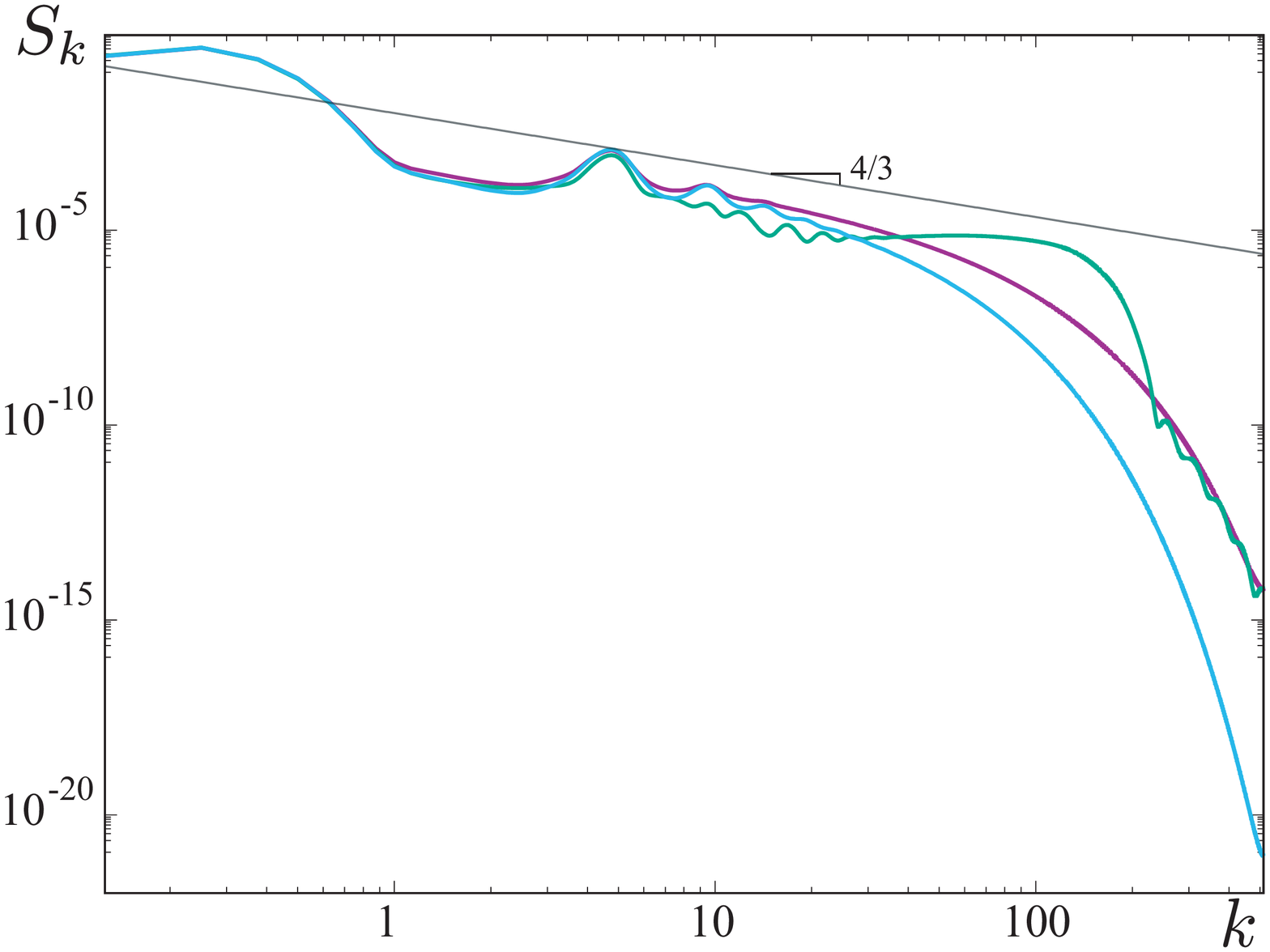}
\caption{
(a) Time evolution of the maximum of the square modulus of the wave-function $|\psi_0(t)|^2$, solution of (\ref{eq:NLS}) with zero forcing and starting at $t=0$ with a smooth initial condition $\psi(x,0)=1+0.1 {\rm cos}(\frac{2\pi  x}{L})$ where $L=51.2$ is the size of the periodic box, $g=\alpha = 1$, for various $\nu$. 
The dynamics shows a peak reminiscent of the self-similar singular dynamics of the inviscid case (the peak is located earlier and its amplitude is larger as $\nu$ decreases). The mesh size has been adapted in order to solve 
the high density peaks and the case $\nu=2.5\times 10^{-4}$ has been slightly shifted in time for the sake of clarity.
(b) Plot of the instantaneous 1D average spectrum $S_k$
as function of  $k$
for the times near the peak for the smallest viscosity $\nu=2.5\times 10^{-6}$. From bottom to top, the blue curve is the available time-step just before, $t=2.546$, the purple one is on the peak $t=2.547$ and the green is for the time-step just after the peak, $t=2.548$. The straight line indicates the slope $k^{-4/3}$ that can be deduced from the self-similar structure of the singularity.}
\label{fig:Turbulence}
\end{figure}

{\it Turbulent behavior.--} Given the dynamics of the NLS equation, it is tempting to investigate the existence of a 
  turbulent regime. 
In this turbulent regime,  mass ($N$) and energy ($H$) are injected at large scale by a forcing term $f_{k_0}({ x},t)$, while the viscous term dissipates at small scales. For practical purposes we characterize the random forcing by  $\left< f_{k_0} \right>=0$, its amplitude $a$ in the Fourier space and its typical scale $k_0$ (See Supplement Materials~\cite{SupplMat}). 
To keep close analogy with NS turbulence, the quantity of interest here will be the mass (\ref{eq:Masse}), positive defined as is the kinetic energy for NS. Following  the NLS equation (\ref{eq:NLS}), the time variation of $N$ reads:
   \begin{eqnarray}
\frac{d N}{dt} = - 2 \nu \int |\Delta\psi|^2 \, d^D{\bm x} +i  \int \left( \psi  f^*_{k_0} -  \psi^* f_{k_0}  \right)  \, d^D{\bm x}.
    \label{eq:Ndissipation}
\end{eqnarray}
Therefore, in strong analogy with fluids, the dissipation (first term in (\ref{eq:Ndissipation})) is strictly negative while the forcing can be positive or negative. On the contrary, time derivatives of eqn. (\ref{eq:Energy}) does not display a simple form as (\ref{eq:Ndissipation}). Thus, in what follows we will focus on $N(t)$ and its dissipation balance (\ref{eq:Ndissipation}).

\begin{figure}[h!]
 \centering 
(a)\includegraphics[width=0.20\textwidth]{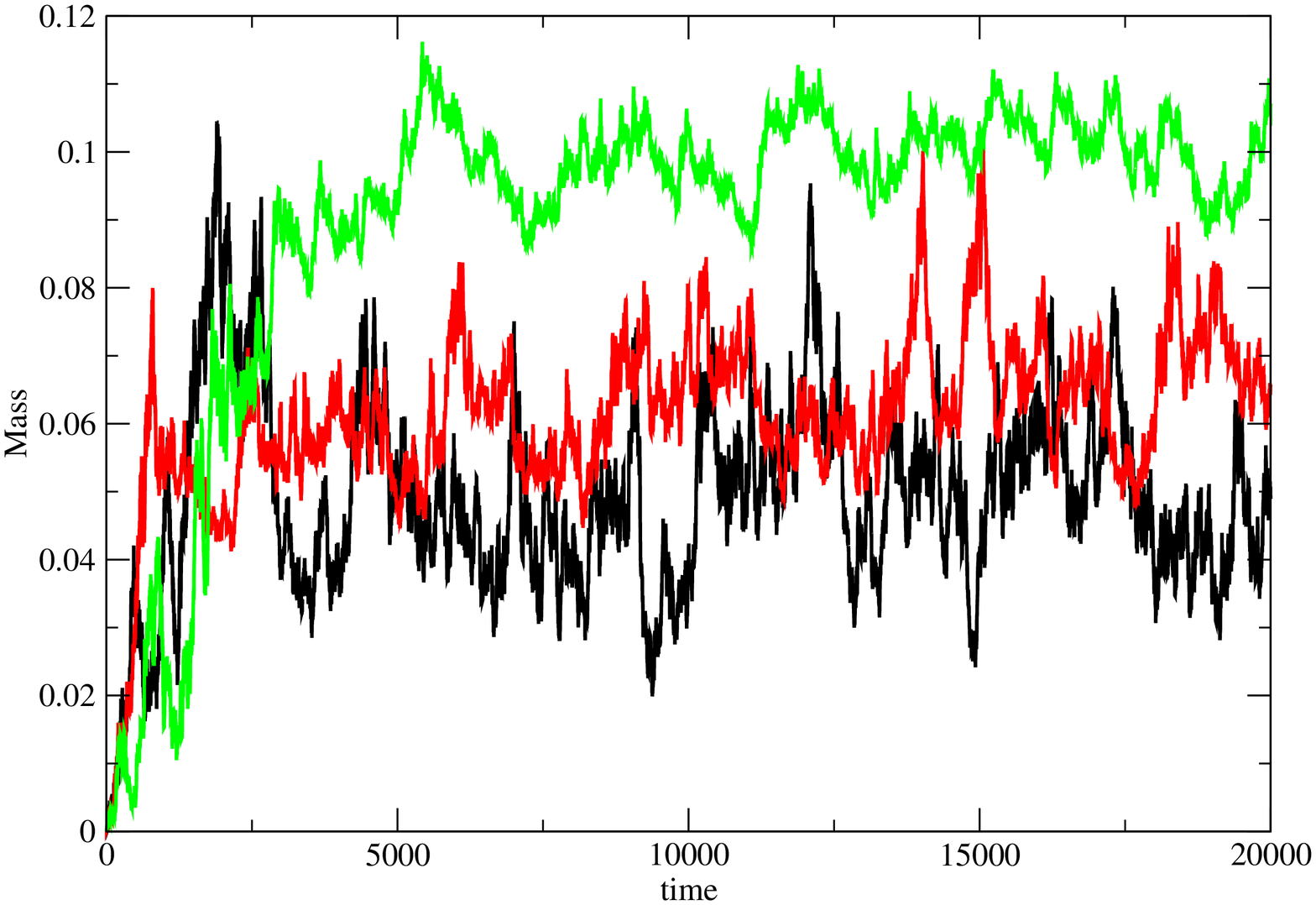} (b)\includegraphics[width=0.20\textwidth]{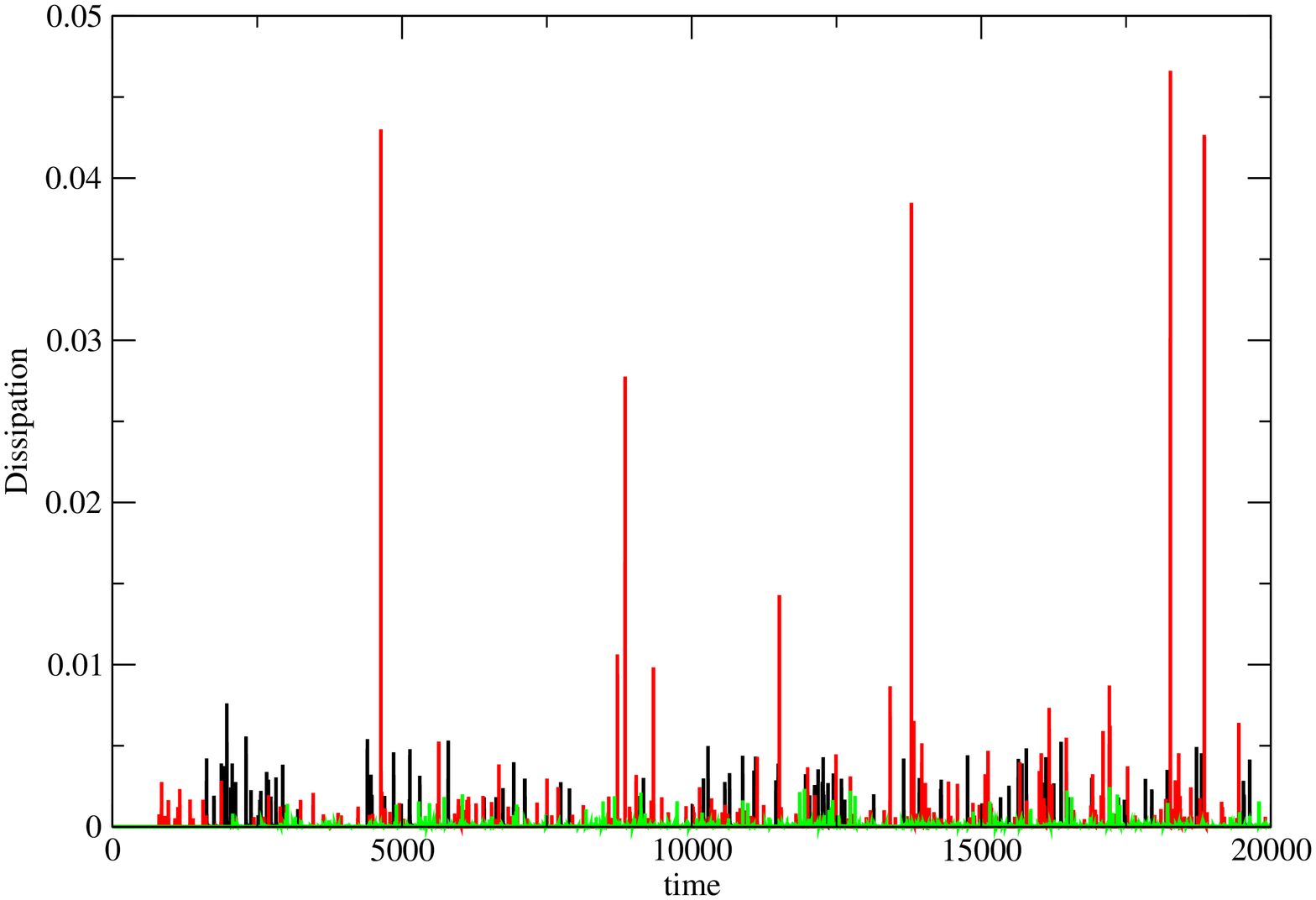}\\
\caption{
 (a) Time evolution of the mean density $\bar N(t)$ for three different values of $\nu$; and, (b) the mean mass dissipation $\bar\epsilon(t)$ is also computed and plotted as function on time for the same values of the viscosity. The numerics is for the same parameters as Fig. \ref{fig:xtDiagram}  but  for $L=307.2$, $dx=0.15$. We have checked that higher resolutions in space and time do not change qualitatively the results.}
  \label{fig:Turbulence2}
\end{figure}

Numerical simulations show the existence of a permanent turbulent regime for some values of the forcing as illustrated on Fig. \ref{fig:Turbulence2}. 
More precisely, it shows the (space) averaged density~\footnote{Throughout the paper, we denote the spatial average of a quantity $q$ by $\bar q$, and the temporal averages by $\left<q\right>$, so that  $\left<\bar\epsilon \right>$ denotes the spatio-temporal average.}  $\bar N(t) = \frac{1}{L}\int_0^L |\psi|^2 dx$ and dissipation $\bar\epsilon(t)=  2 \nu \frac{1}{L}\int_0^L |\partial_{xx} \psi|^2 dx$ as function of time, for three different values of the viscosity $\nu $.
Fig. \ref{fig:Turbulence2} (a) shows that a stationary statistical regime is rapidly reached where $\bar N(t)$ fluctuates around a mean value that is not depending strongly and more particularly not monotonously on the viscosity. The mean (in space) dissipation, Fig. \ref{fig:Turbulence2} (b), exhibits a (statistically steady) randomly distributed sequence of peaks, in close correspondence with turbulent dissipation \cite{Sreenivasan87}. In our picture, these peaks correspond to the formation of singularities that are cured by the viscosity. Remarkably, the dissipation peak decreases with the viscosity, while the peak frequency increases with it: this is not contradictory since $\bar\epsilon(t)$ involves the product of $\nu$ with $ |\partial_{xx}  \psi|^2$, this latter increasing when $\nu$ decreases, but the behavior of the product $\nu  |\partial_{xx}  \psi|^2$ is not prescribed {\it a priori} and  the numerics shows in fact that it decrease as $\nu\to 0$.
  
\begin{figure}[h!]
\centering 
 (a)\, \includegraphics[width=0.2\textwidth]{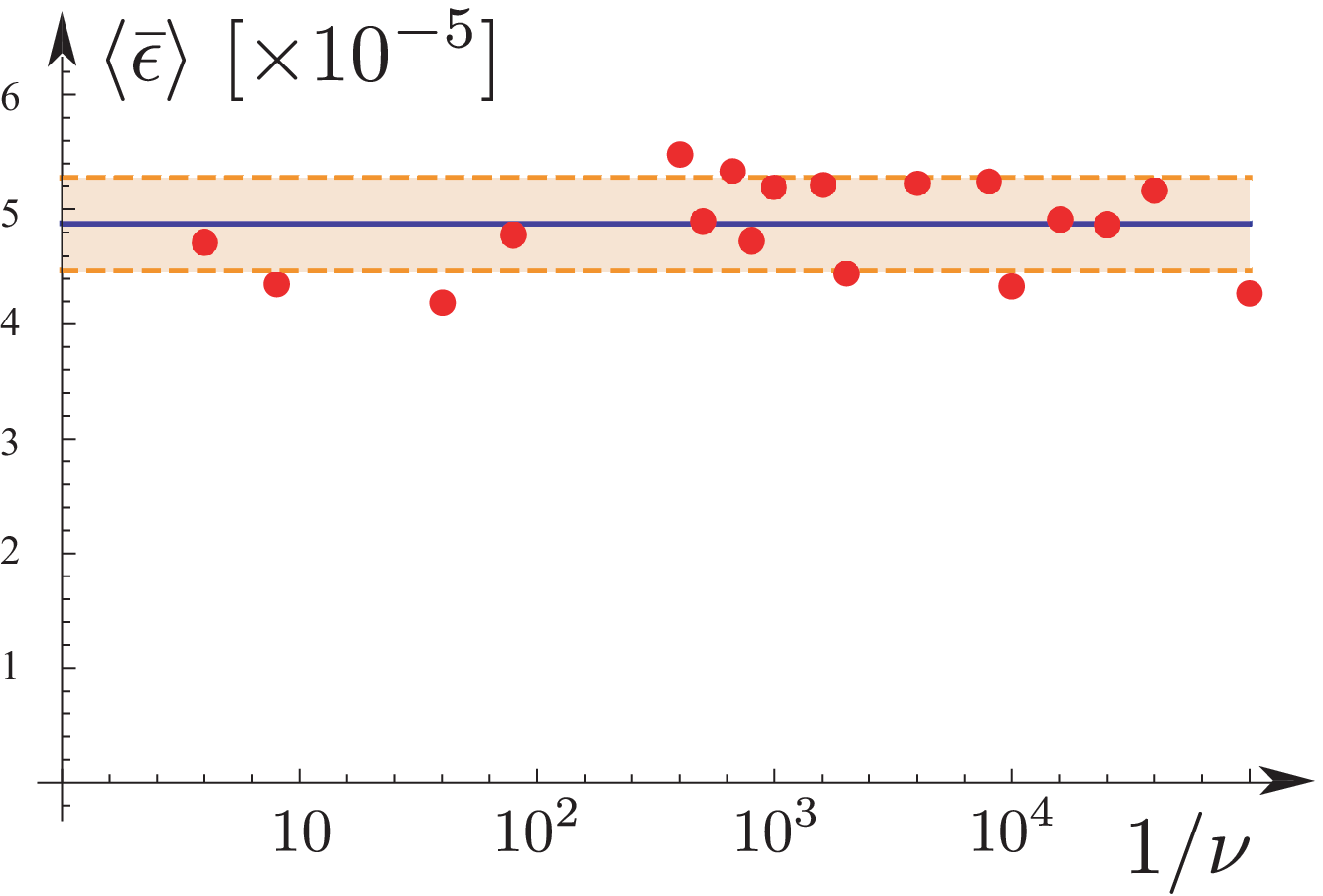}\,\,
 (b)\, \includegraphics[width=0.2\textwidth]{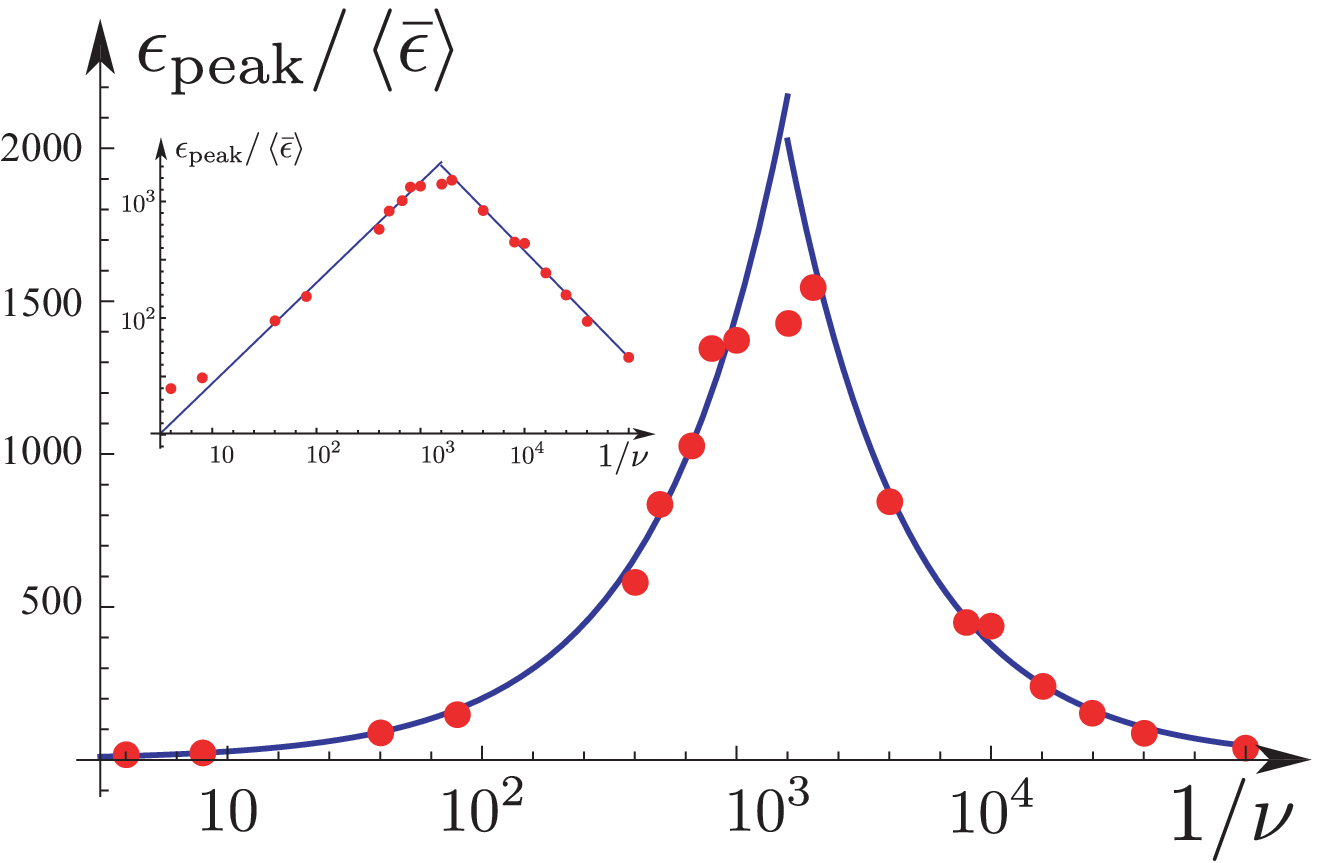}\\
 \caption{(a) Mean dissipation rate $\left<\bar \epsilon \right> $ as a function of viscosity. One notices that the mean dissipation rate is $\left<\bar \epsilon\right> = (4.87 \pm 0.40)\times 10^{-5}$ and almost independent of viscosity. (b) Mean peak dissipation created by the intermittent events, normalized by the mean dissipation rate, as a function of viscosity. One observes that the peak events may be as large as 2000 the mean dissipation rate. The inset suggests a power law $\epsilon_{\rm peak}/\left<\bar \epsilon\right>  \sim A_\pm  |\nu_c/\nu|^{\eta_\pm}$, with $\eta_+ \approx -0.91 $ for $\nu_c/\nu >1 $, $\eta_- \approx 0.86 $ for $\nu_c/\nu<1 $, and $\nu_c\approx 6.56\times 10^{-4}$. }
  \label{fig:Dissipation}
\end{figure}

As the viscosity $\nu$ varies, keeping all the other parameters constant, we observe that the overall picture is preserved. In particular, the mean dissipation rate, obtained by taking the temporal averaged of the space averaged dissipation rate, namely $\left<\bar\epsilon \right>$,
is only slightly varying with the viscosity, as shown on Fig. \ref{fig:Dissipation} (a).
It indicates that the injection determines this quantity and that the dissipation process adapts automatically as $\nu$ varies. It corresponds thus to the ideal situation expected in developed turbulence of anomalous dissipation where it converges to a constant value as the viscosity (an analog of the Reynolds number for fluids) decreases~\cite{Sreenivasan84}. 
On the other hand, the amplitude of the 
dissipation peaks exhibits a non-monotonic dependence with the viscosity as shown in Fig. \ref{fig:Dissipation}(b) where the average of the dissipation peaks normalized by the mean dissipation rate $\left<\bar\epsilon \right>$ is plotted as a function of $1/\nu$. In particular, it is observed that as the viscosity diminishes, the dissipation peaks decrease, similar to the observations of Fig. \ref{fig:Turbulence2} (b).

{\it Kolmogorov spectrum.--}
In the current scenario, eqn. (\ref{eq:NLS}) displays a turbulent behavior in the sense that the injected mass at large scales is eventually dissipated at
the small ones. However, as seen above the dissipation mechanism is far from a cascade regime since it depends from random peaks reminiscent of the singular behavior of the inviscid model. It is thus interesting to measure the spectrum (\ref{eq:SpectrumDef}) that will be obtained for this intermittent spatio-temporal turbulence. 
The spectrum $S_k$ evolution is ruled  by the general transport equation (in Fourier space):
$ \frac{\partial S_k }{\partial t} = - \frac{\partial Q_k }{\partial k} -2\nu k^4 S_k  + F_k, 
$
that relates the flux $Q_k$, the dissipation, $-2\nu k^4 S_k$, and the injection, $F_k =  Im [ 2i  \psi_{\bm k} \bar f_{-{\bm k}}]$. Dissipation acts mostly on the small scales (large $k$ such that $\nu k^4 \gg \alpha k^2$) while injection acts only here on the large scales ($|k|<k_0$) and this scale separation defines an inertial window where dissipation and injection are negligible. 
Similarly to what has been observed for the spectrum variation near a peak (see Fig. \ref{fig:Turbulence}(b)) and because of the existence of intermittent collapses, the instantaneous spectrum $S_k(t)$ fluctuates significantly with time at short scales while the large scales part of the spectrum is roughly independent of time. It is thus more relevant to investigate the averaged (in time) spectrum $\langle S_k \rangle$ in the statistically steady regime, that will, in fact, play the role 
of the Kolmogorov spectrum in fluid turbulence. Thus
 $ \langle  S_k \rangle$ is time-independent and the averaged flux $\langle Q_k \rangle$ becomes constant in the inertial window in $k$. Furthermore, the behaviors at large and small $k$ impose $\langle Q_k \rangle = -2\nu \int_k^\infty k^4 \langle S_k \rangle dk \equiv -\left<\bar \epsilon\right>$. 


\begin{figure}[h!]
\includegraphics[width=0.35\textwidth]{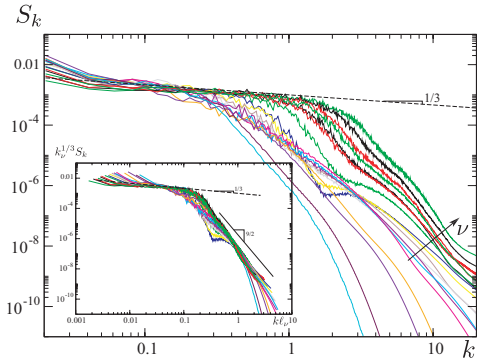}
\caption{ Time averaged spectra $\langle S_k \rangle$ for 18 different values of the viscosities, ranging from  $\nu=2.5 \times 10^{-6}$ up to  $\nu=0.25$. The spectra seem to follow the same scaling law $\langle S_k \rangle \sim k^{-1/3}$  over a range that goes from the injection scale towards a scale that depends on $\nu$. (Inset) The rescaled  spectra $\ell_{\nu}^{-1/3} \langle S_k \rangle$ as function of the rescaled wave number $k \ell_{\nu}$ where $\ell_{\nu}\sim   \left( {\nu}/{\bar\epsilon} \right)^{{3}/{14}}$ is the Kolmogorov dissipation scale. The dashed and continuous line indicate the $k^{-1/3}$ and $k^{-9/2}$ behavior respectively.  }
  \label{Fig:Spectre}
\end{figure}

Fig. \ref{Fig:Spectre} displays such Kolmogorov spectra obtained from direct numerical simulations of (\ref{eq:NLS}) with varying $\nu$, the other parameters being the same as for the preceding Figs. \ref{fig:Turbulence2} and \ref{fig:Dissipation}. We observe that these spectra are almost similar for small $k$ while their extension in $k$ increases as the viscosity decreases. It draws an asymptotic master curve that exhibits a scaling consistent with $\langle  S_k \rangle \sim k^{-1/3}$. As $\nu$ decreases the size of the inertial range increases while the spectrum follows a single master curve. Remarkably, when the amplitude of the forcing $a$ is varied at fixed viscosity, changing, therefore, $\left< \bar \epsilon\right>$, we observe the same independence of the spectra with $\left< \bar \epsilon\right>$.

 A Kolmogorov-like argument is not sufficient to predict such a spectrum, because eqn. (\ref{eq:NLS}) has two distinct conservation laws, leading to an extra dimensionless number (the same situation happens in the case of zero thickness elastic plates \cite{DuringPRF}).
By dimensional analysis and according to the numerical observation that the final spectrum does not depend  on $\left<\bar \epsilon\right>$ one concludes  that  the spectrum $ \langle S_k \rangle \sim \left({\alpha}/{g^{4}}\right)^{1/3}k^{-1/3}$. Remarkably, this final argument is different than for wave turbulence so that our spectrum is not at all of this type~\cite{NazBook}. Moreover, it is different than that of the singularity observed on fig. \ref{fig:Turbulence} suggesting a subtle average of the singularity spectrum, that could be similar to the one proposed in vortex
bursting~\cite{Lundgren82,Maurel03}.

\begin{figure}[h!]
\includegraphics[width=0.35\textwidth]{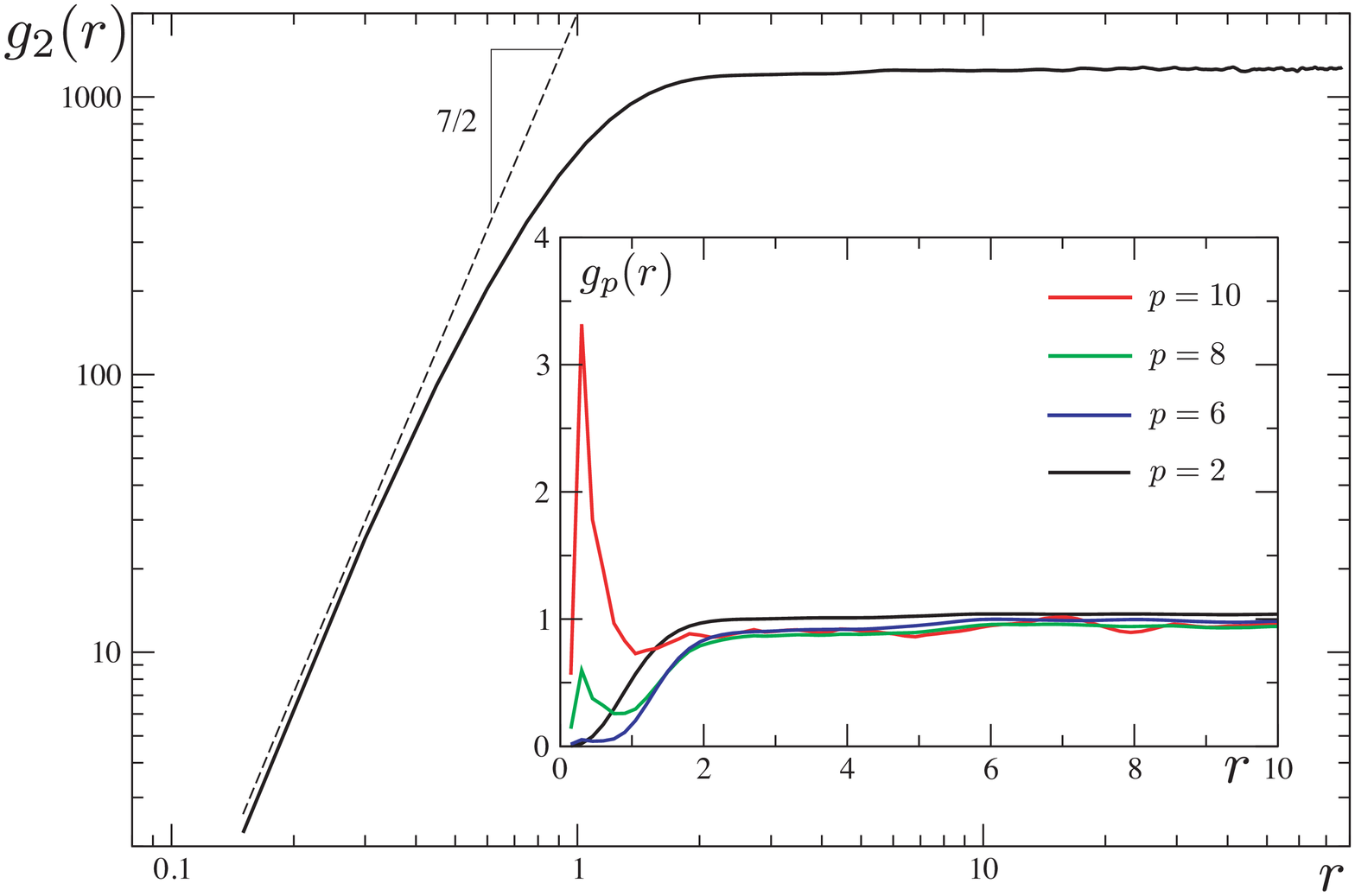} 
\caption{Structure function $g_p(r)$ as function of $r$ for $\nu=10^{-5}$: the main figure shows in a log-log plot the function $g_2$ where
the short scale behavior $g_2(r) \propto r^{7/2}$ is reminiscent of the viscous behavior of the spectrum $S_k \propto k^{-9/2}$.
The inset shows the function $g_p(r)$ in a linear plot for $p=2,\,6,\, 8$ and $10$, normalized by their asymptotic values reached for large $r$. A peak emerges at small $r$ for $p=8$ and more clearly for $p=10$.}
  \label{fig:Sfunc}
\end{figure}

This spectrum is well defined in the inertial range $k$, until a critical scale $\ell_{\nu}$, the so-called
Kolmogorov scale at which the local dissipation rate balances the mean one:
 $\left< \bar\epsilon\right> \sim \nu |\Delta \psi|^2 \sim \nu \psi^2/\ell^4\sim \nu \left( \frac{\alpha}{g} \right) ^{1/3} { \ell}^{-14/3} $ leading to $\ell_\nu  \sim\left(\left( \frac{g}{\alpha} \right) ^{1/3}  \frac{\bar\epsilon}{\nu}\right)^{-\frac{3}{14}}$. This provides an estimate of the scale extent of the spectrum for our 1D problem. Moreover, our results suggest the following general relation for the spectrum:
\begin{equation} 
 \langle S_k \rangle= \left(\frac{\alpha}{g^{4}}\right)^{1/3}\ell_{\nu}^{1/3} G\left(k \ell_{\nu} \right),
 \label{eq:SelfSpectrum}
\end{equation}
where the function $G(\eta)\sim \eta^{-1/3}$ in the inertial range.
The master scaling relation  (\ref{eq:SelfSpectrum}) is tested on the insert of Fig. \ref{Fig:Spectre} where all rescaled spectra $k_{\nu}^{1/3}\langle S_k \rangle$ are
plotted as function of the rescaled wave number $k/k_{\nu}$ taking $k_\nu=1/ \ell_{\nu}$ for different values of the viscosities studied numerically. We observe a reasonable collapse of the spectra for the smallest values of 
the viscosities, showing the self-similar structure of the turbulence.
Moreover, the small scale spectra (large $k$) follow also a
power law behavior that can be fitted by $ G(\eta) \sim \eta ^{-9/2}$. Therefore, the shape of this turbulent spectrum is consistent with a Kolmogorov-type cascade spectrum and does not exhibit the genuine singular dynamics of our equation. Such intermittency of the dynamics should in fact be characterized using
the second order structure functions~\cite{falcon2010,Chibbaro}:
\begin{equation}
 g_p(r)= \overline{ | \psi(x+r)+\psi(x-r)-2\psi(x)|^p}.
 \label{eq:StrucFon}
\end{equation}
These structure functions are particularly sensitive to the high amplitudes of the field at large $p$. From the spectrum we can infer  that the second order structure function $g_2(r) \sim r^{7/2}$ for small $r$ (where dissipation dominates) and 
$g_2(r) \sim r^{-2/3}$ at the integral scale. Figure \ref{fig:Sfunc} shows the structure functions for $p=2,\, 6,\, 8\, \& \, 10$
for a small value of the viscosity $\nu=10^{-5}$. The scaling behavior for $g_2(r)$ is observed at small scale, while the structure function saturates at large $r$, reminiscent of the random fluctuations of the wave-function. More interestingly, the short scale behavior of the structure functions varies abruptly for large $p$ showing a peak that becomes higher as $p$ increases (see inset of Fig. \ref{fig:Sfunc}). We interpret this peak as the signature of  the existence of quasi-singularities events in space and time which create large and narrow density peaks.  Indeed, in principle, the measure of large order structure functions maybe useful to catch singularities in high Reynolds number fluid motion  \cite{martine19}.

 {\it vi) Discussion.-}
We have presented a model-equation with many remarkable properties making it a promising template for investigating the role of intermittencies in real fluid turbulence. The fluctuations of the solutions of equation (\ref{eq:NLS}) display indeed strong analogies with real fluid turbulence. In particular, they exhibit a well-defined Kolmogorov spectrum in an ``inertial range" between the injection at large scales and dissipation at small scales. Moreover, the model shows a phenomenon of strong intermittency that results from the random occurrence of quasi-singularities that are stopped before the final blow-up by a viscous-like term present in the model. This shows well that the occurrence of singularities for the ``inviscid" part of the equation of motion is a way to explain both how dissipation and intermittency occur in such a turbulent system. The flux of a conserved quantity from large to small scales can somehow be linked to the random occurrence of coherent structures, the quasi singularities, with well defined time dependence which is hard -if not impossible- to catch by looking only at space and time-averaged quantities.

SR thanks to FONDECYT grant N$^\circ$ 1181382 and the Gaspard Monge Visiting Professor Program of \'Ecole Polytechnique (France).

\bibliography{wave}
\end{document}